# Negative Effects of Incentivised Viral Campaigns for Activity in Social Networks

Radosław Michalski[1], Jarosław Jankowski[2], Przemysław Kazienko[1]

[1] Institute of Informatics, Wrocław University of Technology, Wrocław, Poland
[2] Faculty of Computer Science, West Pomeranian University of Technology, Szczecin, Poland
radoslaw.michalski@pwr.wroc.pl, jjankowski@wi.zut.edu.pl, kazienko@pwr.wroc.pl

*Abstract*—Viral campaigns are crucial methods for word-of-mouth marketing in social communities. The goal of these campaigns is to encourage people for activity. The problem of incentivised and non-incentivised campaigns is studied in the paper. Based on the data collected within the real social networking site both approaches were compared. The experimental results revealed that a highly motivated campaign not necessarily provides better results due to overlapping effect. Additional studies have shown that the behaviour of individual community members in the campaign based on their service profile can be predicted but the classification accuracy may be limited.

## I. INTRODUCTION

The most important goal of every social network campaign is to maximize the spread of information across the network while keeping the campaign budget on a minimum level. Depending on the type of the campaign, whether it is an informative, activity or sales oriented one, methods to achieve this goal do differ. However, each time when a new social network viral campaign is planned, there is a need to consider whether it is required to support it by incentives of any kind. In the case where it should be stimulated, researchers provide a variety of theoretical models to maximize the effects of incentivised campaigns [1], [2], [3] or scenarios for social networks users rewarding [4], [5]. However, there is still a great need to compare the speed and effectiveness of incentivised viral campaigns on real datasets [6], especially by preserving the same social network environment to compare the obtained results for different strategies. To supply this field with new experimental results and new research ideas, this paper presents such a comparison by using real-life social network data, where two similar viral campaigns were started simultaneously, the first one with incentives supporting activity and the other one without them. Differences between those campaigns' effects are presented, with a special focus on information diffusion speed, effectiveness, user role and his involvement in the campaign. As the results show, incentivised campaigns for activity are not necessarily better than those without incentives. The authors present the conclusions about the possible negative effects of incentivised campaigns. Besides pointing out negative aspects, the novelty of this paper is that authors analyse effects of incentivised campaign after it was finished, what they call *inertia of the incentivised campaign*. All of the dyad types involved in viral campaigns by senders' and recipients' role are defined as well, and those users were later classified to know what group they may fit into while starting a new campaign.

The next section of this paper presents the problem of the effectiveness in viral, word-of-mouth campaigns in more detail by referencing to related work in the subject area. Section four presents the description of the social platform where the two campaigns took place. The main section of this paper, Section IV, presents results of the experiment and the last one, Section V, concludes and presents future work ideas.

## II. PROBLEM DESCRIPTION AND RELATED WORK

Viral campaign management in the online environment includes several stages related to the content design [7], seeding strategies [8], predicting campaign states with mathematical models [9], monitoring [10] and psychological aspects [7].

In the field of psychology, it is known that external incentives are not necessarily the best when considering people motivation [11], [12]. Users often are motivated by themselves (intrinsic motivation), because they want to reciprocate to others [13] or would like to gain higher social approval [14]. In both cases any kind of economic incentives may not stimulate as much as the influencer would like to [11]. Another important factor of motivating people is to ask them to perform some activity interesting from their point of view, but their claims often are more optimistic than their behaviour in comparison to extrinsic incentives [15], [16]. That leads to the conclusion that incentives may be helpful for motivating users to perform some activity, however it is often a problem of how the optimal motivation should look like.

To have better control over the results and campaign reach, marketers can influence campaign dynamics using techniques based on incentives like discounts, referral bonuses or revenue sharing approach [17], [18]. Incentives can affect campaign dynamics positively but also concerns about the evaluation of their effectiveness may be found [19]. For example, when having too many incentives, the range and number of engaged users may go beyond the budget and it may go out of control even after ending the main campaign. With incentives, it is questionable whether the campaign will be properly targeted [20], how many incentives should be given and how large they should be [21]. Content can be delivered to customers that are interested only in incentives and are outside of the target market [22]. Large proportion of messages can be sent only because of incentives thus resulting in the lower quality of infections and potential of marketing action. In traditional approach without incentives users are motivated with personal factors, emotions, social reasons or altruism [7]. As a result content is distributed between users with higher social relations and with potentially better effectiveness. Other concerns are related to repeated messages which are sent to single customers. Earlier research identified that limited number of repetitions can positively affect purchase intentions but above the saturation point which is reached very quickly a drop in effectiveness will be observed [5]. With intensive sending of viral content typical for incentivised campaigns it can be treated as unsolicited messages and cause negative effects [23]. Our research addressed the effectiveness of viral actions and focused on effects of campaigns performed within social networks with and without incentives. The common limitation of earlier research was that it concentrated mainly on independent campaigns without the ability to directly compare different approaches [23], [8]. In this kind of analysis, it was difficult to predict how a campaign would perform with or without incentives. In this paper, the authors analysed two similar campaigns performed in the same conditions and environment with two approaches. The advantage of this research is that it was performed on real life data. Additionally, many earlier researches were performed on simulations and the authors emphasise the need for real data analysis [23]. During the research, it was possible to analyse participants of both campaigns and observe the effects of incentives on the overall campaign reach. The engagement in the campaigns and the characteristics of customers receiving and sending viral content of both types were analysed. In this paper, specifics of both strategies are presented, as well as levels of engagement in both campaigns and characteristics of users affecting the engagement in both campaigns.

### III. Viral Campaigns for Activity

#### A. Social Platform Description

The research was performed within the social platform working in a form of virtual world available in Poland. The system connects functions of chat and entertainment platform, in which users are represented by graphical avatars, and they have the opportunity to engage in the life of the online community. The key features are related to avatars and decorative elements, styles, clothes and virtual products.

During the two week period of the campaigns, it was analysed that 16695 users logged in to the system and 5112 (30.67%) of them engaged in at least one of the two viral campaigns as a receiver or a sender. Based on the demographic of the users who were engaged within this time it was found that 3697 were females (70.36%) and 1515 were males (29.64%). The attributes of the users used in this research are related to measuring the overall user activity represented by *Experience* factor (avg=146.72), total number of logins to the system represented by *Logins* variable (avg=254.37), number of private messages sent and received within the internal communication system denoted respectively as $Msg_{out}$ (avg=493.38) and $Msg_{in}$ (avg=536.99), number of friends with distinguished inbound connections $Friends_{in}$ (avg=215.42) and outbound connections $Friends_{out}$ (avg=144.99). Users are connected with social relation based on connections and friends lists. The main metrics of social network connecting participants of the campaign are represented in Table I.

TABLE I. NETWORK METRICS

| Network metrics | Value |
| --- | --- |
| Nodes | 5,112 |
| Edges | 158,443 |
| Average node degree | 74.33 |
| Network diameter | 6 |
| Average length of shortest paths | 2.462 |

Social connections between users are weighted with parameters related to communication activity and messages sent between users with attributes $Msg_{ab}$ and $Msg_{ba}$. Apart from the global activity where the distinguished parameters representing short term activities within 30 days before the campaign were denoted for messages as $Msg_{out,30}$ (avg=43.15) and received $Msg_{in,30}$ (avg=43.44), $Friends_{in,30}$ (avg=14.77) and $Friends_{out,30}$ (avg=13.58). An additional attribute *Payments* (avg=2.05) is related to the number of transactions and purchased premium services that was used. This attribute is important from the marketing point of view and it is related to the analysis of ARPU (average revenue per user) and ARPPU (average revenue per paying user) as the main parameter of business model evaluation. Other financial factors were based on virtual cash at the user account and was denoted as *Vcash* (avg=212.22). Virtual cash can be received from other users, can be won playing games, or added in a form of bonus after purchasing premium services. Social networking platform and virtual world engine with message systems makes it possible to launch viral campaigns. Viral messaging is a common way of distributing digital objects like gifts or avatars. The target group of internal viral campaigns related to new virtual products or avatars mainly includes users interested in purchasing premium services. In this paper, the viral action was performed during a special event connected with the introduction of two new avatars and two campaigns were conducted in the same period of time. The main aim of this research was to analyse how viral content can spread within the network with different strategies based on non-incentivised and incentivised approach.

*B. Incentivised Campaign A*

Incentivized campaign denoted as A was based on a new avatar related to event. The only way of getting it was by receiving it from other users. After receiving the avatar, the user had ability to send it to friends. Users were motivated to send avatars because of competition with prizes for users who distribute maximal number of avatars and prizes for randomly selected participants. The campaign started with ten randomly selected seeds among logged users. The dataset covered 13 days of objects diffusion within the network but the incentivised part of the campaign ended after ten hours. This kind of execution delivered interesting dataset showing how the campaign was performing during the incentivised period and the inertia effect after it. During the first day 1733 users received the avatar and 354 of them (20.43%) decided to redirect it to friends. The 13 day statistics showed that 3874 users infected and 1873 of them (48.35%) decided to transmit the content to friends. Statistics of the viral connections and messages redirections show the average node degree 1.931, network diameter 16 and average path length 6.258.

*C. Non-incentivised Campaign B*

Non-incentivised campaign B was based on other avatar similar in functionality and style to incentivised one to avoid effects related to design preferences. Transmission to other users was possible in the same way like in campaign A. The viral action was started with randomly selected ten seeds. Users were not receiving any incentives for distribution and all infections were related to natural interest of products. During the campaign, it was observed that there was a push activity when users were recommending avatar and sending to friends and a pull activity from users seeing others wearing new avatar and asking them to send the new product. However, both types of infections were not distinguished and all infections were treated in a same way. Activity within the first day of action shows 1038 infections and 506 (48.75%) of infected users decided to transmit the object to at least one friend. During the whole period of 13 days 3069 users received new object and 1899 of them (61.88%) decided to transmit content to friends. Using campaign data the infections network was built based on users engaged in non-incentive action. The statistics that was obtained show the main network characteristics like average node degree 1.863, network diameter 23 and average path length 6.802.

## IV. RESULTS

*A. Campaign Effects Comparison*

Both campaigns were analysed within a period of thirteen days. For every campaign there were two types of infections analysed – unique and non-unique. Unique infection means that a user received the viral message for the first time, non-unique means receiving multiple messages for the same user. Why is it worth to distinguish and compare both infection types? From the users' point of view he would like to receive the same message only once, because while receiving multiple invitations to perform some activity, after reaching his patience threshold, he may threat this campaign as a spam which should generally not happen, because he may get a negative attitude towards the campaign.

Statistics of both types of campaigns were presented in Table II.

TABLE II. BASIC STATISTICS OF VIRAL CAMPAIGNS

| Property | Incentivized campaign A | Non-incentivised campaign B |
|---|---|---|
| Number of viral messages sent | 28,446 | 9,972 |
| Number of unique senders | 2,886 | 2,683 |
| Number of unique recipients | 3,889 | 3,074 |
| Avg no. of viral msg sent per sender | 9.86 | 3.72 |
| Avg no. of viral msg receiverd per recipient | 7.31 | 3.24 |
| Avg no. of internal messages exchanged between sender and recipient in unique transm. | 0.461 | 0.948 |
| Avg no. of internal messages exchanged between sender and recipient in non-unique transm. | 1.152 | 2.027 |
| Percentace of common senders in both campaigns | 34.3% | |
| Percentace of common receipients in both campaigns | 28.5% | |

For every campaign, we counted the unique and non-unique transmissions during the analysed period grouped in one-hour periods. They are illustrated in Figure 1 (a-d).

At this point, it may be observed that despite the huge amount of viral messages sent in incentivised campaign (almost three times bigger volume of messages), results of information spreading are almost similar – only a 26% increase in the number of unique recipients who received the viral message. All the others (74%) received the message multiple times, despite the fact that they knew it already. The huge peak in Figures 1(a-b) for incentivised campaign is due to the limited time the avatar transmission was offered, as described in Section IV.

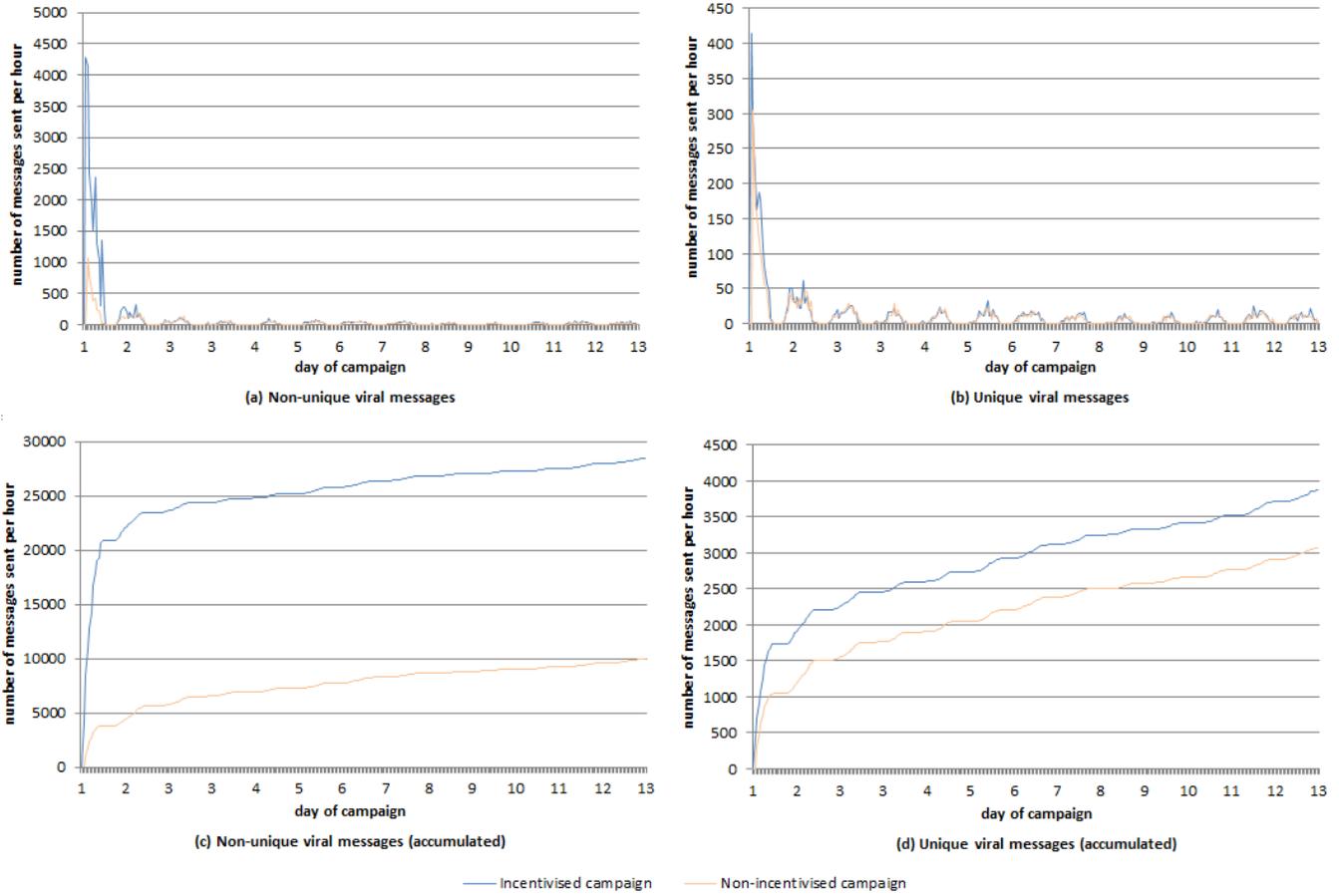

Figure 1. Results of incentivised campaign compared to non-incentivised one

What may be also interesting is the ratio of unique to non-unique messages sent which is presented in Figure 2.

It clearly points out that the effectiveness of incentivised campaign is rather small in comparison to the second one.

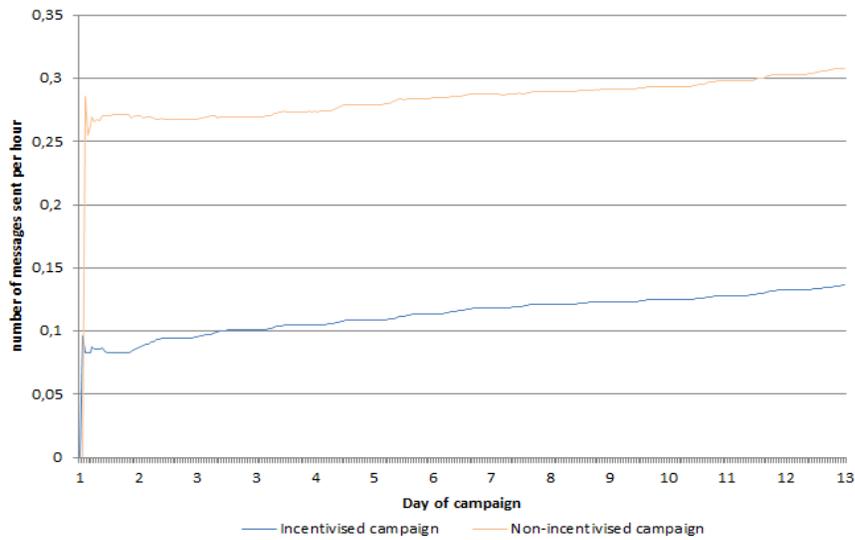

Figure 2. Ratio of unique to non-unique messages in both campaigns

## B. Inertia of the Incentivised Campaign

When analysing the incentivised campaign, it could be seen that the transmission of avatars did not finish after ten hours, but it was still transmitted after the end of the contest. During the contest, 20,917 messages were sent, while 7,529 were sent later and the ratio of users sending the avatars during the contest to those sending after was 1:4. More information about this situation is presented in Table III.

TABLE III. INCENTIVISED CAMPAIGN DURING AND AFTER THE CONTEST

| Property | During the contest | After the contest |
|---|---|---|
| Number of viral messages sent | 20,917 | 7,529 |
| No. of senders sending avatars only in one period | 506 | 2050 |
| No. of senders sending avatars during both periods | 330 | |
| Avg number of viral cash per sender | 453.8 | 199.3 |
| Avg user experience (specific for this portal) | 296.3 | 153.1 |

It may be observed that different groups were involved in sending the viral content during the contest and after it. Less number of users generated huge volume of messages, so they were more focused on winning the contest, but later less than a half of them were interested in sending the viral content further. After going in deeper detail, the difference between some crucial user attributes is also observable – more experienced users were interested only in the contest, which proves that the rest sending the avatar later may not even know the precise rules of the contest, but they simply forwarded the received message without reflecting or checking whether they will benefit from it. This situation was called by the authors' *inertia of the incentivised campaign* – when the viral action still takes place and the consequences of it monitored as well. So it is important to observe what is happening after the initial period due to the fact that some trouble may arise during the campaign. For example, if users notice that they were performing some action without the chance of getting rewarded (because they were not aware of the rules) they may think negatively about the campaign. Therefore, whether the campaign should continue after the competition has ended should be taken into account.

## C. Factors Affecting Receiving Viral Content

For both senders and recipients of viral messages multiple regression models were built with the number of viruses received and sent as dependent variables and set of parameters collected from the system as independent variables. The dataset includes 5112 users participating in the campaign and among them 1448 participating in a passive way (only receiving viral content) and 3664 in an active way (distributing content). In the first stage of the study, the variables that influence the number of infections obtained for both types of virus were identified. The results indicate that, to obtain a virus the greatest importance was related to the parameters associated with the number of incoming contacts in the social network $Friends_{in,\ 30}$ (b = 0.117716, p = 0.000912) and the activity in the frequency of system usage in the period preceding the campaign $Logins_{30}$(b=0.102417, p= 0.000054). The number of received viruses was affected by the *Experience* variable (b=0.095032, p=0.010929) and by the number of transactions made in advance represented by *Payments* (b=0.048614, p=0.014776). The analysis performed for the virus B shows a larger share of social and communication factors among variables affecting the results. The biggest impact on the number of infections was identified on received messages $Msg_{in}$ (b=0.293555, p= 0.000000), $Msg_{in,\ 30}$ (b=0.194994, p = 0.001758), incoming social connections $Friends_{in,30}$ (b=0,196077, p=0.000000) and *Experience* (b=0.192970, p=0.000000). The negative impact on the number of B infections received and messages sent both on a global basis $Msg_{out}$ (b=-0.145451, p= 0.002394) and in the period preceding the campaign $Msg_{out,\ 30}$ (b=-0.139078, p=0, 009 522). This result may indicate that users initiating communications show greater activity when sending messages; in group of lower social status with fewer friends; and when the viral message is addressed to them with less intensity than for users who have a higher position and receive more messages. When $Friends_{in,30}$, $Msg_{in}$ and $Msg_{in,\ 30}$ was increased it was observed that it had a positive effect on the number of infections. This indicates a greater share of the infected hosts in the campaign of B. In summary, for the B virus a greater impact on the number of infections had social factors. Virus B was distributed between the users who are more likely to send messages and have more friends. Campaign B had more in common with real recommendations because of the bond that existed between the sender and recipient.

## D. Factors Affecting Non-unique Sending

In the next step, the authors analysed factors affecting the total number of sent viruses of both types using all messages including multiple but not unique infections. In the campaign A, the greatest impact on the number of infection sent had the user experience *Experience* (b=0.213924, p=0.000000) and frequency of service usage during the preceding campaign period $Logins_{30}$ (b=0.173429, p=0.000000). The number of infections was positively affected by the number of incoming social connections $Friends_{in}$ (b=0.126089, p=0.001040), $Friends_{in,30}$ (b=0.072358, p=0.039134). The interest in the action was influenced by the number of units of virtual currency *Vcash* (b=0.070385, p=0.000016) gathered at a user account which indicates the interest of collecting virtual currency and importance to this form of activity on the site for this group of users. The negative influence of the number of transactions represented by a *Payments* variable (b=-0.075488, p=0.000130) was also highlighted. This indicates less involvement in the distribution of virus "A" users who were making payments and small prizes were not enough to motivate them to join the campaign and to send infections. The results indicate that the target group paying for services was less likely to participate in activities related to the incentives. The negative effect was shown for incoming $Msg_{in,30}$ (b=-0.148785,

p=0.000130), and social connections outgoing $Friends_{out}$ (b=-0.043822, p=0.029471). For the number of B infections sent, the biggest impact was communication activity associated with sending messages $Msg_{out}$ (b=0.266384, p=0.000001) and incoming connections in the structures of social networks in both the shorter horizon $Friends_{in,30}$ (b=0.122330, p=0.000000) and from the beginning of an account creation $Friends_{in}$ (b=0.122330, p=0.001597). In the distribution of media B negative effect of *Payments* and *Vcash* on the user account was not identified. This indicates a greater potential for this type of campaign by reaching the users from the target group than with campaign B.

*E. Factors Affecting Unique Sending*

In the next step, the authors analysed factors affecting engagement in sending unique viruses. The results are presented in the Table IV. For sending virus and unique infections the greatest impact was related to communication activity and messages in a short time horizon $Msg_{out,30}$ (b= 0.240524, p=0.000023) and *Experience* (b=0.231068, p= 0.000000).

TABLE IV. FACTORS AFFECTING UNIQUE SENDING

| Attribute | Campaign A | | Campaign B | |
|---|---|---|---|---|
| | b* | p-value | b* | p-value |
| *Experience* | 0.231068 | 0.000000 | 0.105718 | 0.000216 |
| *Logins* | 0.007407 | 0.843140 | 0.008628 | 0.818711 |
| $Msg_{in}$ | 0.152520 | 0.007948 | -0.410444 | 0.000000 |
| $Msg_{out}$ | -0.124639 | 0.010946 | 0.349491 | 0.000000 |
| $Friends_{in}$ | 0.037784 | 0.391880 | 0.248219 | 0.000000 |
| $Friends_{out}$ | -0.059063 | 0.006637 | -0.132900 | 0.000000 |
| *Payments* | -0.025618 | 0.188812 | -0.041022 | 0.036418 |
| $Logins_{30}$ | 0.047280 | 0.086710 | -0.002336 | 0.932909 |
| $Msg_{in,30}$ | -0.224222 | 0.000955 | 0.022338 | 0.743298 |
| $Msg_{out,30}$ | 0.240524 | 0.000023 | -0.007669 | 0.893056 |
| $Friends_{in,30}$ | 0.049898 | 0.197303 | 0.051534 | 0.185454 |
| $Friends_{out,30}$ | -0.022194 | 0.360718 | 0.084583 | 0.000537 |
| *Vcash* | 0.012923 | 0.416345 | 0.015013 | 0.347756 |
| *Age* | -0.017185 | 0.246322 | 0.019520 | 0.190367 |

The lower level of influence was related to the number of received messages $Msg_{in}$ (b=0.152520, p=0.007948). But in a short-time horizon, the messages $Msg_{in,30}$ (b=-0.224222, p=0.000955) received and sent in the long term $Msg_{out}$ (b=-0.124639, p=0.010946) were affecting results negatively. The negative impact was identified for social network out coming connections $Friends_{out}$ (b=-0.059063, p=-0.059063). For the virus B and unique infection the greatest impact was related to communication activity in the longer term view $Msg_{out}$ (b=0.349491, p=0.000000), the number of incoming connections in the social network $Friends_{in}$ (b=0.248219, p=0.000000). Less important was the *Experience* (b=0.105718, p=0.000216) and outgoing calls in a short-time horizon $Friends_{out,30}$ (b=0.084583, p=0.000537). Infections were negatively affected by the number of unique messages received $Msg_{in}$ (b=-0.410444, p= 0.000000) and out coming social connections in the longer-time horizon $Friends_{out}$(b =-0.132900, p = 0.000000).

*F. Users' Groups Analysis*

During the campaign analysis, the eight groups of users with different levels of involvement in the action of A and B were distinguished. Table V shows the quantitative characteristics for each group and their participation in the campaigns. The number of received infections (Rcv A, Rcv B), total sent viruses (Sent A, Sent B), and unique infections (sent UA, UB sent) is specified. The differences between the groups were compared using the U Mann-Whitney test. The analysis of the members of the groups G1 and G2 shows the important differences with respect to the activity and frequency of use with the variable representing the greatest difference in frequency of use *Logins* Z = 9.84812, p = 0.000000), with an average ranking in the test *Avgrank (G1)* = 1011, 8 and *Avgrank (G2)* = 719.0 in favour of G1. The big difference between users was identified for the group which was designated for incoming calls in a social network $Friends_{in}$ (Z=9.38871, p=0.000000). Larger differences existed in the analysis of variables associated with the activity in the longer term than in short term. A negative correlation was identified only for age, indicating a greater involvement in the action of members of the lower age group. When comparing the groups G3 and G4, no differences were observed at an acceptable level of significance apart from *Payments* with slight difference of the G3 (Z=2.296994, p=0.021620) and the amount of virtual currency associated with the user account (Z=1.889236, p=0.058861).

TABLE V. GROUP ANALYSIS

| Group | Description | Users | Rcv A | Rcv B | Sent A | Sent B | Sent UA | Sent UB |
|---|---|---|---|---|---|---|---|---|
| G1 | Received A & B and sent A & B | 1450 | 16236 | 6551 | 25425 | 7933 | 2823 | 2206 |
| G2 | Received A & B | 437 | 3358 | 1106 | - | - | - | - |
| G3 | Received A & B and sent A | 236 | 2376 | 737 | 1130 | - | 159 | - |
| G4 | Received A & B and sent B | 238 | 1737 | 742 | - | 526 | - | 188 |
| G5 | Received A and sent A | 521 | 2167 | - | 890 | - | 494 | - |
| G6 | Received B and sent B | 314 | - | 419 | - | 440 | - | 305 |
| G7 | Received A | 713 | 2083 | - | - | - | - | - |
| G8 | Received B | 298 | - | 353 | - | - | - | - |

In the next step, the author examined the members of groups G5 and G7. The biggest differences between the groups at a significance level of p <0.05 occur when the *Experience was* (Z=5.409985, p=0.000001), the activity of communication for both $Msg_{in}$ was (Z=4.988272, p=0.000001), $Msg_{out}$ was (Z=4.918889, p=0.000001), networking in a social network and network $Friends_{in}$ was (Z=4.97137, p=0.000001) and $Friends_{out}$ was (Z=4.50486, p=0.000001). For the group G3 there was also more activity regarding *Payments* (Z=4.688503, p=0.000003). The G5 group, therefore, depended on length of service usage and long-term activity. An increase in the age limit decreased the activity in the distribution of virus A as well as the interest in the incentives. The analysis of the groups G6 and G8 indicates that these groups differ on the significance level of p <0.05 values of the attributes associated with *Payments* (Z=2.078155, p=0.037695) and the amount of virtual currency possessed *Vcash* (Z=2.125952, p=0.033507). For both variables, higher values were obtained in the group G6 with engagement in virus B distribution. This indicates a greater interest in new product and distributing it further with users who are making payments and are interested in new services. For groups G1 and G3 the major differences between groups at a low level (Z=2.493669, p=0.012643) for *Friends$_{in}$* occur in the light of incoming connections to lists of friends within the social network, number of *Logins* (Z=2.153197, p=0.031303), *Msg$_{in}$* (Z=2.231340, p=0.025659) and *Msg$_{out}$* (Z=2.354610, p=0.018542). Members who distributed only one content were less engaged in community activities and communication. Another situation is for groups G1 and G4. The biggest difference was observed between the G1 and G4, with the number of payments *Payments* (Z=4.636971, p = 0.000004). G1 and G3 was also largely influenced by the number of links in social network *Friends$_{in}$* (Z=4.197551, p=0.000004). A bigger difference was observed between G1 vs. G4 than between G1 vs. G3.

*G. Classification of Users*

To confirm whether it is possible to predict users' affiliation to a particular group defined in previous subsection (G1-G8), the authors decided to setup an experiment in Weka classification environment [24]. A set of features were used and the general properties of the social network users were: general portal experience, number of logins, number of messages sent and received, number of friends, age, gender, and amount of virtual cash. The following algorithms were evaluated and used for the classification: naive Bayes [25], J48 decision tree [26], decision stump [27], SMO [28], multilayer perceptron [29], and SVM [30]. The 10-fold cross-validation method was the validation method that was used [31]. Initially, the classification results were unacceptable, because the average accuracy was about 35%. In that case, the authors decided to distinguish between the two classes only – G1 and all the others, because the involvement of a user in both campaigns proves that incentives had no or limited influence on his decision to participate in particular campaign. Table VI presents results of classification for the two classes.

TABLE VI. CLASSIFICATION RESULTS FOR TWO CLASSES (G1 VS. ALL OTHERS)

| Classification algorithm | Accuracy (%) | Time (s) |
|---|---|---|
| Naïve Bayes [25] | 68.34 | 0.06 |
| J48 decision tree [26] | 67.63 | 0.45 |
| Decision stump [27] | 65.53 | 0.06 |
| SMO [28] | 69.08 | 2.21 |
| Multilayer perceptron [29] | 69.34 | 23.74 |
| SVM [30] | 69.17 | 7.07 |

V. CONCLUSIONS AND FUTURE WORK

Earlier research related to information diffusion, viral marketing, and models are based mostly on simulations which were criticised by several authors. In this research, the data sample from real campaigns was used and the behaviours observed were based on the social interaction in the real-life online environment.

The experimental results revealed that during the incentivised campaign the less important success factor was the social position and engagement of a user in the community. Non-incentivised campaign engaged users and integrated them more with the community - messages with new products were sent rather to friends. What was more significant was the relationship between the sender and receiver. It was shown that incentivised campaigns are similar to mass media campaigns where messages are broadcasted to groups of untargeted users, even though this may not be necessarily the goal of the campaign initiators.

This kind of less invasive activity can be located between real recommendations among friends and seeding performed by the campaign organiser. Such activity takes on a form of supporting initial seeding with the advantage that it is performed by community members and the responsibility for sending messages is transferred to community members. This way the campaign organiser is not treated as a massive sender.

The percentage of users engaging in non-incentive campaign among all infected was higher than incentivised. It shows that users receiving non-incentivised content were more motivated to forward messages because of emotions, altruism and other sociological factors identified as a key for successful campaign.

Incentives can result with very high commitment in actions confirmed in the research by a high but not unique infection number. Simultaneously, it negatively affects receivers of such messages and causes them to treat messages as a spam. But senders engaged in this kind of action can be negatively affected because incentives were only available to a limited number of winners or the incentive value was too low to attract users with relatively small number of infections. This way high commitment and efforts to collect incentives usually result in low rewards. In the observed campaign, after releasing results of the competition, many users were frustrated with the results – they did not win any prize even spending a lot of time on infecting others. This shows that incentives strategy should be adjusted to the audience profile and many factors should be taken into account during the planning of the campaign.

However, in this research the authors did not have the opportunity to analyse how the value of incentive affects open campaign dynamics and range. The campaigns studied were performed in the closed community with no possibility to spread anything outside. This situation is similar to most social platforms which are natural environments for viral marketing actions and their community is the main target. Another case may occur for large scale campaigns operating without tight boundaries. Despite these limitations, the results of the campaigns revealed that incentives can be treated as an important part of the strategy but they not necessarily positively influence the final range of the campaign. Marketers should be aware that it is hard to predict negative effects of this type of actions, however, some clues about those were provided in this paper.

When about future work directions, authors will try to compare experimental results and theoretical models proposed by others and extend them with parameters related to incentives and other forms of motivation. An interesting area is searching for effective strategies of reaching target groups within community and adapting strategy to social network characteristics and user attributes. A study on differentiated incentives values and their influence on user behaviour in the same controlled environment is planned. Simultaneously, there is a need to verify that similar results are to be found in other social platforms.

ACKNOWLEDGMENT

The work was partially supported by fellowship co-financed by the European Union within the European Social Fund, The Polish Ministry of Science and Higher Education, the research project 2010-13. The calculations were partially carried out in Wroclaw Centre for Networking and Supercomputing (http://www.wcss.wroc.pl), grant No 177.